\documentclass[12pt]{article}
\usepackage{epsfig,amsfonts,amssymb}
\usepackage{hyperref}
\usepackage{cite}
\input epsf.sty
\usepackage{cite}

\def\ZZZ{{\hbox{ Z\kern-1.6mm Z}}}
\def\RRR{{\hbox{ R\kern-2.4mm R}}}
\def\CCC{{\hbox{ C\kern-2.0mm C}}}
\def\zzz{{\hbox{z\kern-1mm z}}}

\newcommand{\qeq}{{\hbox{=\kern-2.3mm ? \kern.5mm }}}
\renewcommand{\qeq}{=}

\newcommand{\OO}{{\cal O}}

\newcommand{\LL}{{\cal L}}

\newcommand{\wt}{\widetilde}
\newcommand{\wh}{\widehat}

\newcommand{\NN}{{\cal N}}

\newcommand{\be}{\begin{equation}}
\newcommand{\ee}{\end{equation}}
\newcommand{\ben}{\begin{eqnarray}\displaystyle}
\newcommand{\een}{\end{eqnarray}}

\newcommand{\bea}[1]{\begin{eqnarray}\label{#1} }
\newcommand{\eea}{\end{eqnarray}}

\newcommand{\refb}[1]{(\ref{#1})}
\newcommand{\p}{\partial}

\def\one{{\hbox{ 1\kern-.8mm l}}}
\def\zero{{\hbox{ 0\kern-1.5mm 0}}}

\begin{document}

\baselineskip 24pt

\begin{center}
{\Large \bf
Entropy Function and $AdS_2/CFT_1$
Correspondence}

\end{center}

\vskip .6cm
\medskip

\vspace*{4.0ex}

\baselineskip=18pt

\centerline{\large \rm   Ashoke Sen }

\vspace*{4.0ex}

\centerline{\large \it Harish-Chandra Research Institute}

\centerline{\large \it  Chhatnag Road, Jhusi,
Allahabad 211019, INDIA}

\vspace*{1.0ex}
\centerline{E-mail:  sen@mri.ernet.in, ashokesen1999@gmail.com}

\vspace*{5.0ex}

\centerline{\bf Abstract} \bigskip

Wald's formula for black hole entropy, applied to extremal black
holes, leads to the entropy function formalism. 
We manipulate the entropy computed
this way to express it as the logarithm of the ground state degeneracy
of a dual quantum mechanical system.
This provides a natural definition of the
extremal black hole entropy in the full quantum theory.
Our analysis also clarifies the relationship between the entropy
function formalism and the Euclidean action formalism.

\vfill \eject

\baselineskip=18pt


Wald's formula\cite{9307038,9312023,9403028,9502009} 
gives an expression for the entropy of a black
hole in terms of the field configurations
near the horizon in any general coordinate
invariant theory of gravity including those with higher derivative terms
in the action. For an extremal black hole whose near horizon geometry
has an $AdS_2$ factor, this formula may be encoded in the entropy
function formlism that reduces the problem of computing the entropy
into a purely algebraic problem for spherically symmetric black 
holes\cite{0506177}
and a problem involving solution of simple differential equations for
rotating black holes\cite{0606244} (see \cite{0708.1270} for a review
and other references). 

The original Wald formula was derived from classical
considerations. 
For various reasons one would like to find
a generalization of this formula in the quantum theory.
In fact once we begin including 
higher derivative corrections to the action in string theory,
the notion of classical action becomes ambiguous since in different
equivalent descriptions of the theory related by duality the classical
actions differ. Thus black hole entropy computed using the
classical action does not respect the duality symmetries of the
theory. Various approaches to addressing this problem have been
suggested, {\it e.g.} using the one particle irreducible (1PI)
effective action instead of the classical action in computing
the 
entropy\cite{0412287,0508042,0510147,0609109,0708.1270}, 
or, in the special case of $\NN\ge 2$ supersymmetric
string theories in four dimensions, the 
OSV formula\cite{0405146,0508174,0601108, 0702146}.

We shall follow a different approach to this problem.
Our goal will be to begin with the expression
for the entropy computed in the classical theory, and relate
it via $AdS/CFT$ 
correspondence\cite{9711200,9802109,9802150} to
the logarithm of the ground state
degenearcy in a dual quantum mechanics living on the
boundary of $AdS_2$. 
The latter can then be regarded as the
definition of the extremal black hole entropy in the full
quantum theory.  Related discussion on $AdS_2/CFT_1$
correspondence can be found in
\cite{9809027,9810251,9812073,9904143,9910076,0008058,
0009185,0412294,0710.2956,0803.3621}.

Before describing the details of our analysis, it will be prudent
to summarize its physical content. The original Wald's formula
for black hole entropy holds for non-extremal black holes, and
in applying this result to extremal black holes we must define the
entropy of an extremal black hole as a limit of the entropy of
a non-extremal black hole. In particular the entropy function
formalism computes the entropy of an extremal black hole only in
this sense. This can be given an alternate
interpretation as follows. If we consider a
black hole that is close to being extremal, we expect its near
horizon geometry to develop a long $AdS_2$
throat, but unlike in the case
of an extremal black hole where the throat is infinitely long, the
throat for a non-extremal black hole will be capped
by a regular horizon.
Neverheless since the throat can be made as long as we want
by going close to extremality, we expect that we should be able
to decouple the asymptotic region from the $AdS_2$ throat
and view the near horizon configuration as a black hole solution
in $AdS_2$\cite{9812073,9904143,0111091,0112056,0412294}. 
This solution, known as the Jackiw-Teitelboim black 
hole\cite{teitel,jack0,jack1},
has been displayed explicitly
in eq.\refb{elor}. The entropy computed by the
entropy function method can be interpreted as the entropy of this
black hole solution in $AdS_2$. 
We find that this agrees with the entropy computed using the
Euclidean action approach\cite{gibhaw}, in agreement with the
general result that  for a  regular black hole
solution the Wald entropy and the one computed
using the Euclidean action formalism
coincide\cite{9307038,0604070}. 
The Euclidean action on the other hand can be related to the
partition function of the quantum mechanics living on the
boundary of $AdS_2$ via the $AdS/CFT$ correspondence. By
studying in detail the temperature dependence of the bulk and
the boundary side of the computation one finds that the
Wald entropy can be interpreted as the logarithm of the ground
state degeneracy of this dual quantum mechanics. The latter can
then be taken as the definition of the entropy of an extremal
black hole even in the full quantum theory.

We begin with a brief review of the entropy function formalism
in a classical theory of gravity.\footnote{For a generic black hole
often the string coupling constant at the horizon is fixed by the
attractor mechanism and cannot be freely adjusted. We are
implicitly assuming that we work with a black hole for which the
string coupling constant at the horizon can be
taken to be small either because it is not fixed or
by adjusting some charges. Only in such cases the classical
approximation makes sense. However 
once we arrive at a statistical interpretation of the classical
Wald entropy using this approximation, we shall use
the statistical entropy as the definition of Wald entropy in the
full quantum theory.}
Extremal black holes will be defined to be those with an
$AdS_2$ factor in the near horizon 
geometry\cite{0705.4214,0803.2998}. 
More precisely
the near horizon geometry of an extremal black hole will have
the structure of a compact space $K$, containing the compact
directions of the theory and also the angular coordinates of 
space-time, fibered over an $AdS_2$ space labelled by the time
coordinate $t$ and the radial variable $r$. It will be useful to
regard the theory in this background as a two dimensional
theory obtained as a result of compactifying the fundamental
theory on $K$\cite{0606244}.
We can then describe the dynamics in the near horizon
geometry of the black hole by
a theory of gravity coupled to a set of 
abelian gauge fields $A^{(I)}_\mu$ and
a set of neutral scalar
fields $\{\phi_s\}$, integrating out all other fields.
Let $\LL_0$ be the classical Lagrangian density and $\Gamma_0$
be the classical action describing the dynamics of these massless fields: 
\be \label{e1pi}
\Gamma_0[g_{\mu\nu}, \{A_\mu^{(I)}\},
\{\phi_s\}]=\int d^2 x\, \sqrt{-\det g} \, \LL_0\, .
\ee 
In this theory we consider a general 
field configuration consistent with the 
$SO(2,1)$ isometry of $AdS_2$. This is of the
form:
\be \label{e0}
ds^2  = v\left(-r^2 dt^2+{dr^2\over 
r^2}\right)   
, \qquad \phi_s =u_s, \qquad  F^{(I)}_{rt} = e^I  \, , 
\ee
where $F^{(I)}_{\mu\nu}=
\p_\mu A^{(I)}_\nu - \p_\nu A^{(I)}_\mu$ and
$v$,
$\{u_s\}$ and $\{e^I\}$ are constants labelling the
background. 
Note that there are no parameters 
explicitly labelling the magnetic charges; 
they are encoded in 
the components of the gauge field strengths
along the
compact directions and appear as discrete parameters labelling the
two dimensional theory. 
We now define:
\be \label{e1}
f(\vec u, v, \vec e)\equiv
\sqrt{-\det g} \, \LL_0=v\, \LL_0\,
\ee
evaluated in the background \refb{e0}. 
Then the black hole entropy is given by
\be \label{e5}
S_{BH}(\vec q) = 2\, 
\pi  \left(
e^I \, q_I - f(\vec u, \vec v, \vec e)\right) \, 
\ee
at 
\be \label{e3}
{\p f \over \p u_s}=0, \qquad {\p
f \over \p v}=0\,,  \qquad {\p f \over \p e^I}=q_I\, .
\ee
The first two sets of equations in \refb{e3} follow from
equations of motion whereas the last equation follows from the
definition of electric charge\cite{0506177}.

Let us now make an analytic continuation $t\to -i\tau$ to 
express \refb{e0} as a solution in Euclidean 
space-time.
We get
\be \label{eec1}
ds^2  = v\left(r^2 d\tau^2+{dr^2\over 
r^2}\right)   
, \qquad \phi_s =u_s, \qquad  F^{(I)}_{r\tau} = -i \, e^I  \, .
\ee
At the next step we introduce new coordinates $(\eta,\theta)$
through the following series of transformations:
\be \label{eseries}
z= \tau + i \, r^{-1}, \quad w= (1+iz)/(1-iz), \quad
\tanh(\eta/2) e^{i\theta} = w\, .
\ee
The complex coordinate $z\equiv \tau + i \, r^{-1}$ describes 
the Euclidean $AdS_2$ as an upper half plane,
and the $SL(2,\RRR)$ isometries of $AdS_2$ act on $z$
as fractonal linear transformations. In the $w$ coordinate system
the upper half plane is mapped into the
interior of a unit disk. Finally $(\tanh{1\over 2}\eta,\theta)$ 
are the usual polar coordinates on the unit disk in the $w$-plane. 
In the
$(\eta,\theta)$ coordinates the
solution \refb{eec1} appears as
\be \label{es2nn}
ds^2  \equiv g^E_{\mu\nu} dy^\mu dy^\nu 
= v\left(
d\eta^2 + \sinh^2\eta \, d\theta^2\right)
\qquad \phi_s =u_s, \qquad  F^{(I)}_{\theta\eta} = 
i \, e^I \, \sinh\eta \, .
\ee
We now note that under the analytic continuation
$\theta \to i\wt t$ and coordinate change $\wt r=\cosh\eta$,
\refb{es2nn} becomes
\be \label{elor}
ds^2 = v \left[-(\wt r^2 -1) d\wt t^2 + (\wt r^2-1)^{-1} d\wt r^2
\right], \qquad \phi_s = u_s, \qquad 
F^{(I)}_{\wt r \wt t} = e^I\, .
\ee
This can be thought of as a black hole solution in $AdS_2$
space with regular horizons at 
$\wt r=\pm 1$\cite{teitel,jack0,jack1}. This is in fact the
solution that we get if we take a black hole close to extremality
and examine its throat region\cite{9812073,9904143}.
To see this we can take a Reissner-Nordstrom
metric
\be \label{em1}
ds^2  = - (1 - a/\rho) (1 - b/\rho) d\tau^2   + {d\rho^2\over
 (1 -a/\rho) (1 - b/\rho)}   
 + \rho^2 (d\theta^2 + \sin^2\theta d\phi^2)\, , \ee
 and consider the limit $\lambda\to 0$ at fixed 
 $(\wt r, \wt t, a)$ with
 \be \label{em2}
 \rho= \lambda\, \wt r+{a+b\over 2}, \quad
 b = a-2\lambda, \quad \tau = a^2 \wt t /\lambda
  \, .
  \ee
  In this limit the $(\wt r, \wt t)$ part of the metric \refb{em1}
  reduces to \refb{elor} with $v=a^2$. In contrast if we had taken
  the extremal limit $a=b$ first and then taken the near horizon
  limit, we would have arrived at the metric given in 
  \refb{e0}\cite{0708.1270}. 
 
We now return to our analysis of the solution
\refb{es2nn} which can be regarded as the Euclidean
continuation of the black hole solution \refb{elor}.
The boundary of the unit disk in the $w$ plane is
at $\eta=\infty$, but
we shall regulate the volume of $AdS_2$ by 
putting an upper cut-off $\eta_{max}$ on $\eta$.  
Thus we take $(\eta,\theta)$ to lie in the range
\be \label{erange}
0\le\eta\le \eta_{max}, \quad 0\le\theta<2\pi,
\quad (\eta,\theta) \equiv (\eta, \theta+2\pi)\, .
\ee
Let us evaluate the classical action in this background.
First note that the analytic continuation
$t\to -i\tau$ gives an extra factor of $-i$ and replaces 
$\sqrt{-\det g}$ by $\sqrt{\det g}$.
Since $\LL_0$ is a scalar it remains unchanged under
a coordinate transformation, \i.e.\ the
value of $\LL_0$ evaluated in the background \refb{es2nn}
is the same as that evaluated in the background \refb{e0}.
According
to \refb{e1}, \refb{e5} this is given by 
$\left(e^I q_I -(2\pi)^{-1} S_{BH}(\vec q)\right)/v$ .
$\sqrt{\det g} \, d^2 x$ is also invariant under a general
coordinate transformation. 
Thus on-shell the action is given by
\ben \label{esk2}
\Gamma_{0} &=&-i \, \int d\eta d\theta\, \sqrt{\det g^E}\, \LL_0
= -2\pi\, i\, v\, \int_0^{\eta_{max}}\, 
d\eta \, \sinh  \eta \,   \LL_0   \nonumber \\
&=& -2\pi \, i \, (\cosh\eta_{max} - 1)
\, f= 
 i \, (\cosh\eta_{max} - 1)
\, (S_{BH}(\vec q) - 2\pi e^I \, q_I) \, .
\een

This is however not the complete contribution to 
$\Gamma_0$; 
we can get additional contribution from the boundary terms 
at $\eta=\eta_{max}$. To determine the
form of the boundary contribution, we make a
change of coordinates
\be \label{ech1}
\wt\eta = \eta_{max}-\eta, \qquad \wt\theta = {1\over 2}
\, e^{\eta_{max}}\, 
\theta\, .
\ee
In these coordinates $\wt\theta$ labelling the coordinate along
the boundary
has period 
\be \label{eperiod}
\beta = \pi\, e^{\eta_{max}}\, . 
\ee
Furthermore the metric and the gauge field strengths
near the boundary take the form
\ben \label{ech2}
ds^2 &=& v\left[ d\wt\eta^2 + 
\left( e^{-\wt\eta} - e^{\wt\eta-2\eta_{max}}\right)^2\, 
d\wt\theta^2\right]
=  v\left[ d\wt\eta^2 + e^{-2\wt\eta}d\wt\theta^2\right]
+ \OO\left(\beta^{-2}\right )\, , \nonumber \\
F^{(I)}_{\wt\eta\wt\theta} &=&  i\, e^I\, 
\left( e^{-\wt\eta} - e^{\wt\eta-2\eta_{max}}\right)
= i \, e^I\, e^{-\wt\eta}  +
\OO\left(\beta^{-2}\right )\, .
\een
Now the boundary term in the action is given by some local
expression involving the various fields integrated along the
boundary. Due to translation symmetry along $\wt\theta$, the
integration along the boundary gives a factor of $\beta$ multiplying
the integrand. On the other hand the form of the solution
given in \refb{ech2} shows that the integrand is given by a
$\beta$-independent term plus a contribution of order
$\beta^{-2}$.
Thus up to correction
terms of order
$\beta^{-1}$, 
the boundary contribution must be
proportional to the length $\beta$ of the boundary
circle.\footnote{This
line of argument is similar to the one used in \cite{0506176} in the
context of $AdS_3$ space.}
Together with \refb{esk2} this gives 
\be \label{esk4}
\Gamma_0 = - i \left[
S_{BH}(\vec q) - 2\pi e^I q_I  + \beta \, 
K(\vec q) +\OO(\beta^{-1})\right]\, ,
\ee
for some constant $K(\vec q)$. Conventionally the terms linear
in $\beta$ are removed by adjusting the 
boundary terms\cite{0506176}, but we
shall not need to worry about them.
One point to note is that the $\beta$ independent
term in $i\Gamma_0$ is precisely $S_{BH}(\vec q)
-2\pi q_I e^I$ without any additional 
normalization factor; this will be important in
what follows.

Let us compare this with the Euclidean action 
formalism\cite{gibhaw}. According to this the action
$\Gamma_0$ is related to the energy $E$, 
entropy $S_{BH}$, charges
$q_I$ and the chemical potential $\mu^I\equiv 
i\ointop d\wt\theta
\left. A_{\wt\theta}^{(I)}\right|_{\eta_{max}}$ via 
the relations:\footnote{Note that our definition
\refb{e3} of the electric charge is such that a point charge
$q_I$ will induce a term $q_I \int A^{(I)}_\mu dx^\mu$
in the action.}
\be \label{eac1}
i\Gamma_0= S_{BH} -\beta E + \mu^I q_I\, .
\ee
Now for the classical background fields \refb{es2nn} we have
\be \label{ekk1}
\ointop d\wt\theta A^{(I)}_{\wt\theta}\bigg|_{\eta_{max}}
= -\int_{\eta\le \eta_{max}} \, d\eta \, d\theta \, 
F^{(I)}_{\theta\eta}
=- 2\pi i \, e^I\, (\cosh\eta_{max}-1) = - i e^I \left(\beta - 2\pi 
+\OO(\beta^{-1})\right)\, .
\ee
This gives
\be \label{eac2}
i\Gamma_0 =
S_{BH}(\vec q) - 2\pi e^I q_I  - \beta \, (E - e^I q_I)\, .
\ee
This agrees with \refb{esk4} in the $\beta\to\infty$ limit
for the choice  
\be \label{echh1}
E(\vec q) = -K(\vec q) +  e^I q_I\, .
\ee
This
is in accordance
with the general result that for a regular black hole the Wald entropy
and the one computed using Euclidean action 
formalism agree\cite{9307038,0604070}.
Earlier  exploration of the direct relation between the
entropy function formalism and the Euclidean action formalism
can be found in \cite{0704.0955,0704.1405}.

We now return to our main goal, which is to give an
interpretation of the entropy $S_{BH}$ appearing in
\refb{esk4} in terms of an appropriate conformal quantum 
mechanics
living at the boundary of $AdS_2$. For this we
recall that $e^{i \Gamma_0}$ is the classical partition
function of the theory on $AdS_2$.\footnote{Note that 
we have used the normalization
and sign conventions appropriate for Lorentzian signature 
space-time.
However the explicit $-i$ factor in the expression \refb{esk2}
for $\Gamma_0$ reflects that we are carrying
out the path integral after analytic continuation to Euclidean
signature space-time.} 
Since we are working in the approximation where the theory
in the bulk is treated classically, 
one would expect this to be
the partition function of the dual
quantum mechanics living at the boundary $\eta=\eta_{max}$.
There is  however one additional point we should remember. 
The usual rules of $AdS/CFT$ 
correspondence\cite{9802109,9802150} tells us that for
every gauge field $A_\mu^{(I)}$ in the bulk we have a 
conserved charge $Q_I$ in the boundary theory.  Furthermore,
in the presence of a non-zero $A_\mu^{(I)}$ field at the boundary,
$e^{i\Gamma_0}$, instead of calculating the partition
function, is actually expected to calculate the expectation value
of $e^{i Q_I \ointop d\wt\theta 
A^{(I)}_{\wt\theta}}$. Now from \refb{ekk1} 
we have\footnote{Note that a change in $\vec e$ not only
induces a change in the boundary value of $\{A_\mu^{(I)}\}$, 
but also induces a change in the values of other fields via the
attractor mechanism. The effect of all these other changes can
be included in the Hamiltonian $H$ of the boundary
theory, thereby making $H$ dependent on $\vec e$. Gauge fields
are special, since besides the $-i\beta e^I$ term in \refb{ekk1} which,
being proportional to $\beta$,
can be included as a correction to $H$, there is a $\beta$ independent
contribution $2\pi i e^I$. This cannot be regarded as a
$\beta$ independent correction
to $H$.}
\be \label{eexp}
e^{i Q_I \ointop d\wt\theta A^{(I)}_{\wt\theta}}
= e^{ Q_I e^I(\beta -2\pi) + \OO(\beta^{-1})}\, .
\ee
Thus if $H$ denotes the Hamiltonian generating
$\wt \theta$ translation\footnote{Since $\wt\theta$ translation
induces a
rotation about the origin in the $w$-plane, $H$ can be identified
with $(L_{-1}+L_1)$ up to an additive constant and a
proportionality factor.} 
in the dual quantum mechanics living at 
$\eta=\eta_{max}$
then, according to $AdS/CFT$ 
correspondence,\footnote{Conventionally one interprets the
location of the boundary as providing an ultraviolet cut-off of
the boundary theory. Here we shall find it more convenient to
regard $\beta$ as providing an infrared cut-off, keeping the ultraviolet
cut-off fixed.}
\be \label{eks1}
e^{i\Gamma_0} = Tr \left(e^{-\beta H + (\beta -
2\pi) Q_I e^I+\OO(\beta^{-1})}\right) =
Tr \left(e^{-\beta H'-2\pi Q_I e^I+\OO(\beta^{-1})}\right)\, ,
\qquad H' \equiv H - e^I\, Q_I\, .
\ee
Since we are working
in the classical limit $Q_I$ can be replaced by the charge $q_I$
carried
by the black hole. Furthermore 
in the $\beta\to \infty$ limit the right  hand side of
\refb{eks1} gets its dominant
contribution from the ground states of $H'$. 
If $d(\vec q)$ denotes the
degeneracy of ground states of the $CFT_1$
then in this limit eq.\refb{eks1} takes the form
\be \label{eks2pre}
e^{i\Gamma_0} = e^{-\beta E'}\, 
d(\vec q) \, e^{-2\pi q_I e^I}\, .
\ee
Comparing this with \refb{esk4}, or equivalently
\refb{eac2},  and noting that $E'=E - e^I q_I$ as a consequence
of \refb{eks1}, we see that the microscopic and the macroscopic
results agree if we identify
\be \label{eks4}
e^{S_{BH}(\vec q)} = d(\vec q) \, .
\ee
Eq.\refb{eks4} gives an interpretation of the Wald entropy
$S_{BH}(\vec q)$ computed using the classical entropy function
as the statistical entropy of the dual $CFT_1$.\footnote{Even
though it appears that $d(\vec q)$ counts only the ground state
degeneracies of the $CFT_1$ we recall that we are working in units
in which the ultraviolet cut-off has been taken to be of order 1.
Thus all finite energy states in this unit are actually infinite
energy states in the conventional unit in which the ultraviolet
mass cut-off is taken to infinity. As a result ground states constitute the
complete set of states of the $CFT_1$.}

In the full quantum theory $e^{i\Gamma_0}$ should be replaced by
path integral over various fields of the theory in the 
$AdS_2$ background.
In order to properly define this path integral
we need to fix the boundary condition on various fields
at $r=r_0$.
 In $AdS_{d+1}$ for general $d$ the classical
Maxwell equations for a gauge
field near the boundary has two independent solutions.
One of these represent the constant mode of the
asymptotic 
gauge field and 
the other one measures the asymptotic electric field or
equivalently the charge carried by the
solution. 
Requiring the absence of singularity in the interior of $AdS_{d+1}$
gives a relation between the two coefficients\cite{9802109,9802150}. 
Thus in defining the
path integral over $AdS_{d+1}$ we fix one of the coefficients and
allow the other one to fluctuate.  For $d\ge 3$ the constant mode of the
gauge field is dominant near the boundary; hence it is natural to fix
this and allow the mode measuring the
charge to be determined dynamically in the classical
limit and to fluctuate in the full quantum theory. 
If we continue to define the partition function this way even
for $d=1$ then
the dual $CFT_1$ will contain states of different charges.
Let us denote by $Z(\beta,\vec e)$  the full quantum
partition function with the boundary condition
\be \label{edefei}
 -{i\over 2\pi}\,  \ointop d\theta \, (A^{(I)}_\theta
-\p_\eta A^{(I)}_\theta)= e^I \, ,
\ee
as in
\refb{ekk1}.
Note the
subtraction term proportional to $\p_\eta A^{(I)}_\theta$; it
removes the piece linear in $\beta$ from 
$\ointop d\theta \, A^{(I)}_\theta$.
Equating $Z(\beta,\vec e)$ to the right hand side of
\refb{eks1} in the
$\beta\to\infty$ limit we get\footnote{A similar equation was obtained
in \cite{0608021} using $AdS_3/CFT_2$ 
correspondence. The analysis
however depended heavily on supersymmetry, and the procedure
used there
to deal with the divergent terms is quite different from ours.}
\be \label{etra2}
Z(\beta, \vec e) = e^{-\beta E'} \sum_{\vec q} \, d(\vec q) \, 
e^{- 2\pi q_I e^I}\, ,
\ee
where $d(\vec q)$ is the degeneracy of ground states carrying charge
$\vec q$ in $CFT_1$.
To see how this gives us back 
\refb{eks4} in the semiclassical limit, we note that
in this limit the left hand side is given by
$e^{i\Gamma_0}$. Also in this limit $d(\vec q)$ is large, 
and the
dominant contribution to the sum comes from the maximum
of the summand as a function of $\vec q$. This gives
\be \label{enm1}
e^I = {1\over 2\pi}\, 
{\p \ln d(\vec q)\over \p q^I}, \qquad
i\Gamma_0 = - \beta E' + \ln d(\vec q) - 2\pi q_I e^I\, ,
\ee
and
\be \label{ens1}
e^{i\Gamma_0} = e^{-\beta E'}\, 
d(\vec q) \, e^{-2\pi q_I e^I}\, .
\ee
Eq.\refb{ens1} gives us back \refb{eks2pre} and 
hence \refb{eks4}, 
whereas \refb{enm1}
recovers the relation between $\vec e$ and $\vec q$ given in
\refb{e3}.

We shall now argue however that $Z(\beta,\vec e)$ defined
above is not
the natural definition of the partition function in $AdS_2$. This
is due to the fact that on $AdS_2$
the mode
that measures the electric field (and hence the charge) 
is the dominant one near the boundary.
This can be seen for example in \refb{ekk1} 
where the term proportional
to $\beta$ measures the electric field 
and the constant term 
is determined in terms of the electric field by requiring
the gauge fields to be non-singular at the origin. 
Hence fixing the constant mode of the gauge field at the boundary
and letting the electric field (and hence the charge) to fluctuate
amounts to integrating over non-normalizable modes. While 
such a definition 
of the 
partition function may work under certain circumstances,  it
is not guaranteed to work in general. Instead
it is more natural to  fix the asymptotic electric fields
and let the
constant mode of the gauge field fluctuate. Since fixing the
asymptotic electric fields fixes the charges, the dual $CFT_1$ will
now only contain states of fixed charge, equal to that carried by the
black hole.
If we denote the $AdS_2$ partition function defined
this way by $\wh Z(\beta, \vec q)$, then \refb{etra2} is
replaced by
\be \label{etra2new}
\wh Z(\beta, \vec q) = e^{-\beta E'} \, d(\vec q) \, 
e^{- 2\pi q_I \langle e^I\rangle}\, ,
\ee
where
$e^{- 2\pi q_I \langle e^I\rangle}$
is the expectation value
of $\exp[iq_I\ointop d\theta \, (A^{(I)}_\theta
-\p_\eta A^{(I)}_\theta)]$ at $\eta=\eta_{max}$. 
Now
for large $r_0$
the subtraction term proportional to $\p_\eta A^{(I)}_\theta$
gives a fixed contribution of the form $e^{C' \beta}$ for some
constant $C'$. 
This leads to the simple formula:
\be \label{ettex}
d(\vec q) = \left\langle \exp\left[-
i  q_I\ointop d\theta \, A^{(I)}_\theta\right]
\right\rangle^{finite}_{AdS_2}\, ,
\ee
where $\langle ~\rangle_{AdS_2}$ 
denotes the unnormalized path integral
over various fields on $AdS_2$ with fixed
asymptotic values of the electric fields, and the superscript
$finite$ denotes that we need to pick the constant
multiplicative factor of this expression removing all
terms of the form $e^{C_0\beta}$ for some constant $C_0$.
In the classical limit 
$\wh Z$ reduces to $e^{i\Gamma_0}$, and 
$\langle e^I\rangle=e^I$
where $e^I$ and $q^I$ are related by the attractor equation
\refb{e3}. Thus we again recover \refb{eks4}.

Eq.\refb{etra2new} can be used to compute $d(\vec q)$ in terms
of the partition function of the appropriate string theory on
$AdS_2$. 
It is then
natural to define the entropy of the extremal black hole
in the full quantum theory as $\ln d(\vec q)$. 
This can then be compared with the statistical entropy
computed using the microscopic description of the black hole.

Finally we would like to point
out that for primitive charge vectors
the quantum entropy defined this way refers
strictly to the entropy associated with single centered black holes
and will not capture the entropies of multi-centered black holes
or any other
configuration with the same total charge. This in particular
will imply that this entropy will not suffer from wall crossing or the
problems associated with entropy enigma\cite{0702146}.

\medskip

\noindent {\bf Acknowledgement:} I would like to thank Justin David,
Rajesh Gopakumar and Dileep Jatkar for useful discussions. I would
also like to thank the people of India for their generous 
support to research in string theory.


\end{document}